\documentclass[onecolumn]{aastex631}

\usepackage{subfigure}

\newcommand{\MgII}{\ion{Mg}{2}}
\newcommand{\FeII}{\ion{Fe}{2}}
\newcommand{\CaII}{\ion{Ca}{2}}

\newcommand{\swift}{Swift}

\begin{document}

\title{A Pilot Survey of an M Dwarf Flare Star with Swift's UV Grism}

\email{shashank.chavali@colorado.edu}
\author[0000-0001-9615-7181]{Shashank Chavali}
\affiliation{Laboratory for Atmospheric and Space Physics, University of Colorado, 600 UCB, Boulder, CO 80309, USA}

\author[0000-0002-1176-3391]{Allison Youngblood}
\affiliation{Exoplanets and Stellar Astrophysics Laboratory, NASA Goddard Space Flight Center, Greenbelt, MD 20771, USA}

\author[0000-0002-8090-3570]{Rishi R. Paudel}
\affiliation{University of Maryland, Baltimore County, Baltimore, MD 21250, USA}
\affiliation{Exoplanets and Stellar Astrophysics Laboratory, NASA Goddard Space Flight Center, Greenbelt, MD 20771, USA}
\affiliation{Center for Research and Exploration in Space Science and Technology, NASA/GSFC, Greenbelt, MD 20771, USA}

\author[0000-0001-5646-6668]{R. O. Parke Loyd}
\affiliation{Eureka Scientific, Oakland, CA 94602, USA}

\author[0000-0002-0502-0428]{Karan Molaverdikhani}
\affiliation{Universit\"ats-Sternwarte, Ludwig-Maximilians-Universit\"at
M\"unchen, Scheinerstrasse 1, 81679 M\"unchen, Germany}
\affiliation{Exzellenzcluster Origins, Boltzmannstrasse 2, 85748 Garching, Germany}
\affiliation{Max-Planck-Institut f\"ur Astronomie, K\"onigstuhl 17, 69117 Heidelberg, Germany}

\author[0000-0002-4489-0135]{J. Sebastian Pineda}
\affiliation{Laboratory for Atmospheric and Space Physics, University of Colorado, 600 UCB, Boulder, CO 80309, USA}

\author[0000-0001-7139-2724]{Thomas Barclay}
\affiliation{University of Maryland, Baltimore County, Baltimore, MD 21250, USA}
\affiliation{Exoplanets and Stellar Astrophysics Laboratory, NASA Goddard Space Flight Center, Greenbelt, MD 20771, USA}
\affiliation{Center for Research and Exploration in Space Science and Technology, NASA/GSFC, Greenbelt, MD 20771, USA}

\author[0000-0002-5928-2685]{Laura D. Vega}
\affiliation{University of Maryland, College Park, MD 20742, USA}
\affiliation{Exoplanets and Stellar Astrophysics Laboratory, NASA Goddard Space Flight Center, Greenbelt, MD 20771, USA}
\affiliation{Center for Research and Exploration in Space Science and Technology, NASA/GSFC, Greenbelt, MD 20771, USA}

\begin{abstract}

The near-ultraviolet (NUV) spectral region is a useful diagnostic for stellar flare physics and assessing the energy environment of young exoplanets, especially as relates to prebiotic chemistry. We conducted a pilot NUV spectroscopic flare survey of the young M dwarf AU Mic with the Neil Gehrels Swift Observatory’s UltraViolet and Optical Telescope. We detected four flares and three other epochs of significantly elevated count rates during the 9.6 hours of total exposure time, consistent with a NUV flare rate of $\sim$0.5 hour$^{-1}$. The largest flare we observed released a minimum energy of 6$\times$10$^{33}$ erg between 1730-5000 \AA. All flares had durations longer than the $\sim$14-17 minute duration of each \swift\ visit, making measuring total flare energy and duration infeasible. 


\end{abstract}

\section{Introduction} \label{sec:Introduction}


The near-ultraviolet (NUV; $\sim$1700-3200 \AA) is a historically under-studied region for stellar flares, but has an enormous potential for diagnosing flare physics. In flare models such as \citet{Kowalski2013}, the broad-band visible light from a flare comes from a sunspot-sized patch of hot, $\sim$10,000 K plasma on the stellar photosphere, making NUV measurements potentially more efficient at constraining this flaring mechanism than any other wavelength range. The NUV is also thought to be an essential energy source for triggering abiogenesis, the origin of life, and open questions remain about whether the cool photospheres of M dwarfs could provide enough NUV photons to initiate chemical reactions that ultimately lead to the production of RNA precursors and whether flares could make up for the deficit in NUV energy \citep{Ranjan2017, Rimmer2018}.



  \begin{figure}
         \centering
          \includegraphics[width=\textwidth]{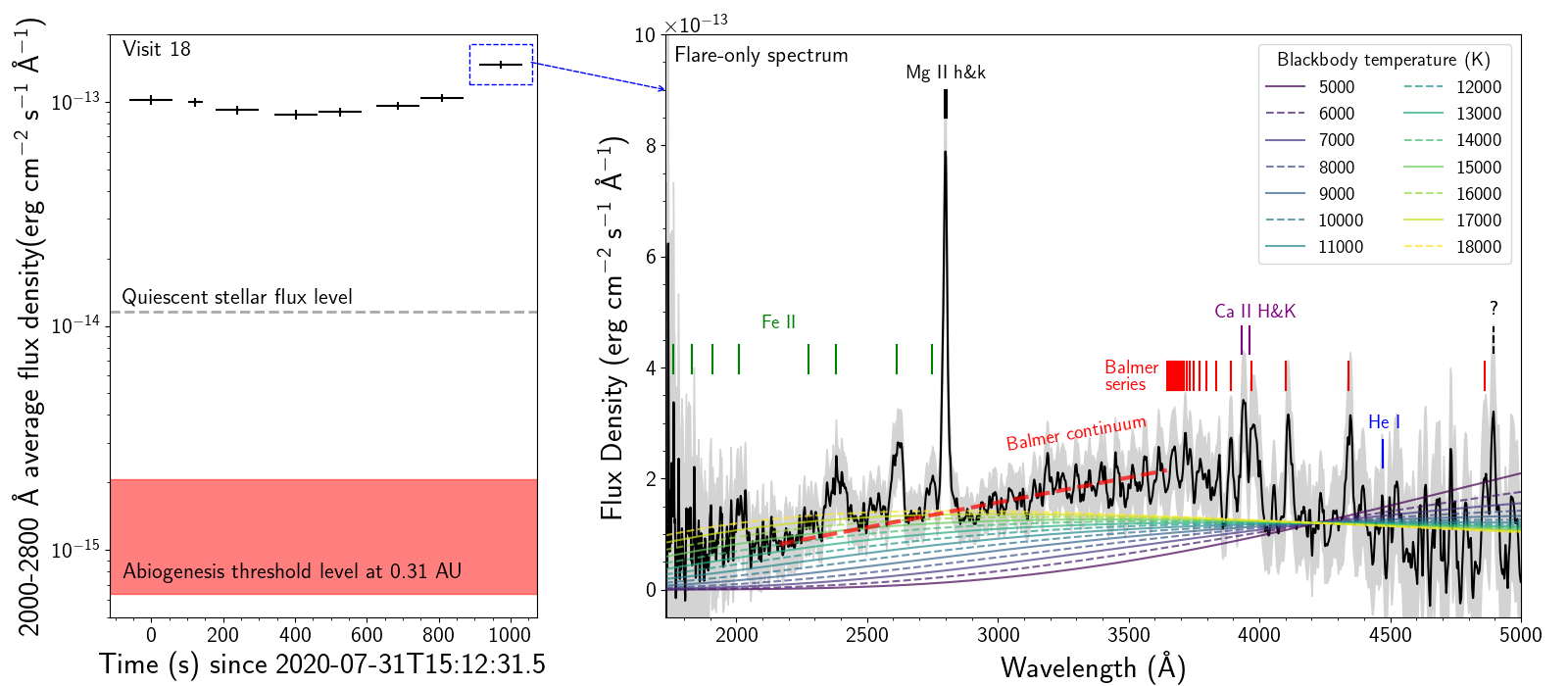}

          \caption{\swift\ grism data from the brightest flaring epoch show that AU Mic regularly exceeds the abiogenesis threshold and that the flaring emission is comprised of chromospheric emission lines, Balmer continuum, and possibly blackbody continuum. Left: the average stellar flux from 2000-2800 \AA\ is shown as a function of time from the start of Visit 18. The red shaded area shows the abiogenesis threshold level from \cite{Rimmer2018} at 0.31 AU (AU Mic's conservative inner habitable zone edge), and the dashed grey line shows the quiescent stellar flux level. Right: a flare-only spectrum (i.e., the quiescent stellar spectrum is subtracted from the in-flare spectrum) of the brightest data point from the left panel is shown as a black line with the gray shading representing the 1-$\sigma$ uncertainty. Various emission lines and the Balmer continuum are labeled. The colored lines show blackbody curves with temperatures ranging from 5000-18,000 K normalized to match the spectrum at 4200 Å. }
          \label{fig:flares}
  \end{figure}

\section{Observations \& Reductions} \label{sec:ObservationsReductions}

We designed a pilot program to determine the viability of NUV spectroscopic monitoring of flare stars with the Neil Gehrels Swift Observatory (\swift) UltraViolet Optical Telescope (UVOT) grism. The UVOT UV grism is optimized for 1700-2900 \AA\ first-order spectra at spectral resolving power $R\sim$150. We targeted AU Mic, a pre-main sequence M0Ve star at 9.7 pc, for its brightness (V=8.6 mag), high flare rate ($\sim$0.5 hour$^{-1}$ based on archival UVOT UVM2 photometry; E. Gilbert et al., 2022 in preparation), and sky location away from the Galactic center where source confusion and oversubscription of \swift\ observing time are less of a concern. We observed AU Mic for 52 ks (14.4 hours) over 42 individual spacecraft visits between 2020 April 5 and 2021 April 2 using the UVOT 0x0384 mode, which alternates between the UV grism (UGRISM, clocked mode, 120~s exposures) and UVM2 images (30~s exposures, used for aspect correction, 1997-2495 \AA), resulting in $\sim$14-18 minutes elapsed time per visit. After overhead, the total target exposure time was 34.5 ks (9.6 hours). Event mode is generally not used with the grism because of the burden on telemetry. 

We reduced the UVOT data using standard procedures from \texttt{HEASoft} \citep{Heasoft} and \texttt{uvotpy} \citep{Kuin2014}. We corrected for offsets in the \texttt{uvotpy} wavelength solutions by fitting the blended Mg II h\&k emission lines with a Gaussian and shifting the centroid to 2800 \AA\ \citep{Kuin2015}. We corrected for contamination by background stars by visually inspecting every 2D spectrum and using linear interpolation to splice out parts of the spectrum contaminated by background stars. These corrections were different for every spectrum because the roll angle of the telescope varies over the course of a year.



\section{Results} \label{sec:Results}
We established median flux levels for the ensemble UVM2 (11.4 counts/s) and grism data and searched for data that lay $>$3$\sigma$ above the median in order to identify flares. In the total 34.5 ks spent on target, we detected four flares (one shown in Figure~\ref{fig:flares}) and three additional epochs where the count rate was significantly elevated above the median, but the light curve morphology was inconsistent with flaring. Note, however, that the cadence and total duration of our visits do not allow us to rule out flaring for these three elevated-flux epochs. Without event mode, 30-120~s cadence combined with 14-18 minute visit durations is a challenging setup for fully resolving flare light curves. We could not determine the start and stop times of any of our flares, so we adopt the visit durations (16.6-18.2 minutes) as the minimum flare durations. For the three elevated epochs, the visit durations were only 14.2 minutes. 

The dates corresponding to the four flaring epochs are: 2020 May 10, 2020 July 05, 2020 July 31 (Figure~\ref{fig:flares}), and 2020 October 18. The three periods of elevated flux that could not confidently be identified as flares occurred on 2020 April 26, 2020 June 28, and 2020 July 26 . The observed flare rate of 0.42-0.73 hour$^{-1}$ is consistent with that expected from archival UVM2 photometry of AU Mic (E. Gilbert et al., 2022 in preparation).


Sufficient NUV photons may be a necessary ingredient for triggering photoreactions that kick off prebiotic chemical pathways. We compared all spectra to the 2000-2800 \AA\ abiogenesis threshold from \cite{Rimmer2018} to determine how flares could impact the ability of M dwarf stars to meet this flux threshold. Unlike older, later type M dwarfs, AU Mic provides $\sim4\times$ the abiogenesis threshold energy during quiescence to its conservative inner-habitable zone edge (0.31 AU; \cite{Kane2022}). Our results indicate that AU Mic spends 98\% of the time $>$5$\times$ above the threshold, 26\% $>10\times$ above the threshold, 4\% $>25\times$ above the threshold, and 0.3\% $>50\times$ above the threshold. Note that AU Mic's two known Neptune-sized exoplanets at 0.06 and 0.11 AU are closer to the star than the habitable zone \citep{Cale2021}.

We subtracted the median spectrum from each of the flare spectra in order to analyze the type of emission processes responsible for the flux increases we observe during flaring and/or elevated epochs. We obtained useful flare spectra between $\sim$1730-5000 \AA\ for our largest flare (Figure~\ref{fig:flares}), which lasted $>$18.2 minutes and released a minimum of 6$\times$10$^{33}$ erg in the 1730-5000 \AA\ bandpass, with roughly equal energy contribution from emission above and below the Balmer series limit at 3646 \AA. We find Balmer emission from H$\beta$ through the Balmer series limit at 3646 \AA, and \CaII\ H\&K emission. Continuum in the 4000-5000 \AA\ region is roughly flat and consistent with a $\sim$6,000-15,000 K blackbody. Shortward of 3646 \AA, we observe bright Balmer continuum emission that dominates over any hot blackbody component. The emission lines shortward of the Balmer series limit are blended multiplets from \MgII\ and \FeII. This flare spectral energy distribution is consistent with that found for GJ 1243 (M4Ve) by \cite{Kowalski2019}. 



\section{Conclusion} \label{sec:Conclusions}
The goal of this \swift\ pilot survey of AU Mic was to explore the utility of \swift's UV grism mode for stellar flare monitoring. Aside from the Hubble Space Telescope, which has detector protection restrictions in place against flaring M dwarfs, \swift's grism uniquely provides NUV spectroscopy for studying stellar flares. We have shown that although the 0x0384 mode's lack of event mode and short visit duration make determining flare morphology and total duration challenging, flux-calibrated flare spectra are readily measurable. Future \swift\ flare surveys should explore the feasibility of consecutive executions of the 0x0384 mode so that individual visits could extend longer.

\begin{acknowledgments}
The authors thank Paul Kuin for advice in setting up this observing program and using \texttt{uvotpy}. This work was supported by NASA's \swift\ Guest Investigator Program under award 80NSSC20K1112.
\end{acknowledgments}


\bibliography{main.bbl}{}
\bibliographystyle{aasjournal}

\end{document}